\documentclass[aps,pra,12pt]{revtex4-2}
\usepackage{amsmath}
\usepackage{amssymb}
\usepackage[unicode=true,colorlinks=true,citecolor=blue,urlcolor=blue]{hyperref}
\usepackage{graphicx}

\begin{document}
\title{Lorentz-Drude dipoles in the radiative limit and their modeling in finite-difference time-domain methods}

\author{Heming Wang}
\affiliation{Department of Electrical Engineering and Edward L. Ginzton Laboratory, Stanford University, Stanford, California 94305, USA}
\author{Shanhui Fan}
\email{shanhui@stanford.edu}
\affiliation{Department of Electrical Engineering and Edward L. Ginzton Laboratory, Stanford University, Stanford, California 94305, USA}

\begin{abstract}
The Lorentz-Drude model for electric dipoles is a classical framework widely used in the study of dipole dynamics and light-matter interactions.
Here we focus on the behaviors of Lorentz-Drude dipoles when their radiative rate dominates their energy loss.
We show that dipole radiation losses do not count toward phenomenological dipole losses if the driving field is interpreted as the total field at the dipole.
In particular, if the dipole does not contain non-radiative losses, then the Lorentz-Drude damping term should be removed.
This is verified by self-consistent implementations of point dipoles in finite-difference time-domain simulations, which also provide a method to directly compute the transport properties of light when dipoles are present.
\end{abstract}
\maketitle

\section{Introduction}

The Lorentz-Drude model \cite{lorentz1916theory} is a classical model that describes the response of an atom to a driving electric field. The model simplifies an atom as a point electric dipole formed by an electron bound to the nucleus. This model has been widely used in both conceptual discussions \cite{bohren2004absorption} and numerical simulations of light-matter interactions \cite{rakic1998optical, vial2005improved}.

In the Lorentz-Drude model, the damping rate can be incorporated as a phenomenological parameter. In typical textbook discussions of the Lorentz-Drude model, this damping rate is interpreted as the total damping rate, which includes both radiative and non-radiative contributions \cite{siegman1986lasers}. In many practical scenarios such as laser gain mediums, the non-radiative rate dominates the radiative rate. In these cases, the total damping rate can be approximated as the non-radiative rate by dropping the radiative contributions, and the use of the Lorentz-Drude model has been well established.

In recent years, there has been significant interest in exploring emitters in the radiative limit, where the radiative rate dominates over the non-radiative rate. For example, an emitter in the radiative limit can behave as a perfect mirror that completely reflects a single photon. This reflection has been considered for both waveguide geometry \cite{shen2005coherent, chang2007single-photon, lodahl2015interfacing} and 2D materials that interact with plane waves \cite{bettles2016enhanced, shahmoon2017cooperative, rui2020subradiant, sheremet2023waveguide}. However, existing theoretical treatments have not made use of the Lorentz-Drude model.

In this paper, we provide a discussion of the Lorentz-Drude model in the radiative limit, focusing in particular on the issues related to the incorporation of the radiative damping rate. We prove that if the driving field is taken to be the total field that the Lorentz-Drude dipole experiences, including both the external field and the induced self-field, then the damping rate should only include the non-radiative component. In particular, when the non-radiative rate is zero, the phenomenological damping rate in the Lorentz-Drude model should be set to zero. However, the motion of the displacement in the dipole still experiences radiative damping when one couples such a lossless dipole to Maxwell’s equations. As an illustration of the formalism, we show that the complete reflection effect \cite{shen2005coherent, stobinska2009perfect} can be treated using the Lorentz-Drude model in this way. We also show that numerical simulations of Maxwell’s equations together with a point dipole described by the lossless Lorentz-Drude model provide an accurate and direct description of the Purcell effect, i.e. the modification of the spontaneous emission rate when the dipole is embedded in an electromagnetic environment, and moreover allow one to compute the modification of the transport properties of light due to the Purcell effect.

The main results of our paper are related to several recent works. The statement that a phenomenological radiative damping rate is not needed in the radiative limit has been noted in the simulation of a point dipole in a dielectric environment \cite{schelew2017self-consistent} as well as in quantum mechanical studies of the interaction of a two-level system with photonic structures \cite{zhou2024simulating}. However, a succinct theoretical proof has not been previously provided.

The paper is organized as follows.
In Section \ref{sec:Hamiltonian}, we derive the Lorentz-Drude model using a Hamiltonian formalism by considering a charged harmonic oscillator coupled to the electromagnetic field. Our derivation shows that radiation damping terms are not necessary when the driving field is interpreted as the total field at the dipole location. We also reconcile our approach with the formalism in which radiation damping has been explicitly incorporated. As an application of the formalism, we treat the effect of complete reflection of light from a planar surface with uniformly distributed dipoles.
In Section \ref{sec:implement} we apply these results to the finite-difference time-domain simulations of a dipole. We derive the Green function on the Yee grid and show that the dipole can be implemented with lossless Lorentz-Drude materials.
Section \ref{sec:examples} provides simulation examples that accurately reproduce various phenomena associated with dipoles in the radiative limit.
Finally, in Section \ref{sec:conclusion} we conclude.

\section{Lorentz-Drude dipoles in the radiative limit} \label{sec:Hamiltonian}

\subsection{Hamiltonian formalism} \label{ssec:H_theory}

We begin by considering a point charge with mass $m$ and charge $q$, bound by a harmonic restorative force and placed in an environment filled with dielectrics. The total Hamiltonian of the system consists of the field part and the charge part \cite{landau2010classical}:
\begin{equation}
H_\mathrm{total} = H_\mathrm{field} + H_\mathrm{charge}
\label{eq:LD_H}
\end{equation}
\begin{equation}
H_\mathrm{field} = \int \left[\frac{\varepsilon(\mathbf{x})\varepsilon_0}{2}\mathbf{E}(\mathbf{x})\cdot\mathbf{E}(\mathbf{x}) + \frac{1}{2\mu_0} \mathbf{B}(\mathbf{x})\cdot\mathbf{B}(\mathbf{x})\right] \mathrm{d}^3\mathbf{x}
\end{equation}
\begin{equation}
H_\mathrm{charge} = \frac{1}{2m}\left[\mathbf{P} - q \mathbf{A}(\mathbf{x})\right]\cdot\left[\mathbf{P} - q \mathbf{A}(\mathbf{x})\right] + \frac{1}{2}m\omega_0^2 \mathbf{X}\cdot\mathbf{X}
\end{equation}
where $\mathbf{E}(\mathbf{x})$ is the electric field, $\mathbf{B}(\mathbf{x}) = \nabla\times\mathbf{A}(\mathbf{x})$ is the magnetic flux density, $\mathbf{A}(\mathbf{x})$ is the magnetic vector potential, $\mathbf{X}$ is the displacement of the charge, $\mathbf{P}$ is the canonical momentum conjugate to $\mathbf{X}$, $\varepsilon_0$ is the vacuum permittivity, $\mu_0$ is the vacuum permeability, $\varepsilon(\mathbf{x})$ is the relative permittivity, $\omega_0$ is the intrinsic resonance frequency of the dipole, and $\mathbf{x}$ is the spatial coordinate for the field. We have chosen the Hamiltonian gauge where the scalar potential is identically zero \cite{jackson2002lorenz} so that the potential term does not appear in $H_\mathrm{charge}$.
For simplicity, we have considered isotropic materials and an isotropic dipole; generalizations to anisotropic cases should be straightforward by replacing $\varepsilon(\mathbf{x})$ and $m$ with appropriate matrices.
We have also assumed that the materials are lossless and dispersionless, such that $\varepsilon_\mathrm{r}(\mathbf{x})$ is a positive real number everywhere. A frequency-dependent permittivity would require the internal dynamics of the material, which changes the form of $H_\mathrm{field}$ \cite{raman2010photonic}. For the purpose of deriving the equations of motion, the conjugate variables for the field are $\varepsilon(\mathbf{x})\varepsilon_0\mathbf{E}(\mathbf{x})$ and $\mathbf{A}(\mathbf{x})$, and the conjugate variables for the charge are $\mathbf{X}$ and $\mathbf{P}$.

The motion of the charge as described by $H_\mathrm{charge}$ will be influenced by both electric and magnetic fields. Assuming that the atom is located at $\mathbf{x}=\mathbf{0}$, if the extent of the motion is small, we can drop the spatial dependence of $\mathbf{A}$ and replace it with $\mathbf{A}_0 \equiv \mathbf{A}(\mathbf{x} = \mathbf{0})$:
\begin{equation}
H_\mathrm{charge} \approx \frac{1}{2m}\left(\mathbf{P} - q \mathbf{A}_0\right)\cdot\left(\mathbf{P} - q \mathbf{A}_0\right) + \frac{1}{2}m\omega_0^2 \mathbf{X}\cdot\mathbf{X}
\end{equation}
This effectively decouples the magnetic field from the charge. The equations of motion as derived from Eq. (\ref{eq:LD_H}) now read:
\begin{equation}
\frac{\mathrm{d}\mathbf{X}}{\mathrm{d}t} = \frac{1}{m}\left(\mathbf{P} - q \mathbf{A}_0\right)
\end{equation}
\begin{equation}
\frac{\mathrm{d}\mathbf{P}}{\mathrm{d}t} = -m\omega_0^2\mathbf{X}
\end{equation}
with $t$ the time variable. If the dipole moment is defined as $\mathbf{p} = q\mathbf{X}$, then its dynamics becomes
\begin{equation}
\frac{\mathrm{d}^2\mathbf{p}}{\mathrm{d}t^2} + \omega_0^2\mathbf{p} = \frac{q^2}{m}\mathbf{E}_0
\label{eq:LD_no_loss}
\end{equation}
where we defined $\mathbf{E}_0 \equiv \mathbf{E}(\mathbf{x} = \mathbf{0})$ and used $\partial \mathbf{A}/\partial t = -\mathbf{E}$ as derived below. This equation, with a restoring term and a driving term, completely describes the dynamics of the dipole in an electric field. Specifically, we note that no radiation loss appears on the left side of Eq. (\ref{eq:LD_no_loss}) since no energy is lost in the combined system of the dipole and the field.

Equation (\ref{eq:LD_no_loss}) is a simple restatement of Newton's law for the dipole and should be fully self-consistent. However, this formulation is different from many textbook introductions to this topic \cite{siegman1986lasers, jackson2009classical}. A more familiar form of the dynamics reads
\begin{equation}
\frac{\mathrm{d}^2\mathbf{p}}{\mathrm{d}t^2} + \kappa \frac{\mathrm{d}\mathbf{p}}{\mathrm{d}t} + \omega_0^2 \mathbf{p} = \frac{q}{m}\mathbf{F}_\mathrm{ext}
\label{eq:LD_with_loss}
\end{equation}
Here the driving force $\mathbf{F}_\mathrm{ext}$ is the external force and can be replaced by $q\mathbf{E}_\mathrm{ext}$ if this force is entirely due to an external electric field. The decay rate $\kappa$ is usually decomposed as $\kappa = \kappa_\mathrm{rad} + \kappa_\mathrm{nrad}$ \cite{siegman1986lasers}, where $\kappa_\mathrm{rad}$ is the rate of radiative decay (or spontaneous emission in the quantum description) and $\kappa_\mathrm{nrad}$ is the non-radiative decay rate. In the limit of $\kappa_\mathrm{nrad} = 0$, the radiative loss still appears on the left side of Eq. (\ref{eq:LD_with_loss}).

In the following, we reconcile this difference by explicitly differentiating the total field at the dipole [as in Eq. (\ref{eq:LD_no_loss})] and the external field at the dipole [as in Eq. (\ref{eq:LD_with_loss})]. The difference between these fields is the self-field $\mathbf{E}_\mathrm{dipole}$ induced by the dipole. The self-field oscillates with a phase difference compared to the dipole oscillation and results in energy exchange between the dipole and the field. As such, the dipole can lose energy to the field even when Eq. (\ref{eq:LD_no_loss}) does not explicitly contain any dissipation terms.

To understand the self-field in the Hamiltonian framework, we consider the dynamics of the vector potential as derived from the Hamiltonian Eq. (\ref{eq:LD_H}):
\begin{equation}
\varepsilon(\mathbf{x})\varepsilon_0 \frac{\partial \mathbf{E}}{\partial t} = \frac{1}{\mu_0} \nabla\times\nabla\times\mathbf{A} - \frac{q}{m}\left(\mathbf{P} - q\mathbf{A}_0\right) \delta(\mathbf{x})
\end{equation}
\begin{equation}
\frac{\partial \mathbf{A}}{\partial t} = -\mathbf{E}
\end{equation}
These two equations can be combined and results in
\begin{equation}
\varepsilon(\mathbf{x})\varepsilon_0\frac{\partial^2 \mathbf{E}}{\partial t^2} + \frac{1}{\mu_0}\nabla\times\nabla\times \mathbf{E} = -\frac{\mathrm{d}^2\mathbf{p}}{\mathrm{d}t^2} \delta(\mathbf{x})
\label{eq:E_wave}
\end{equation}
This is the wave equation for $\mathbf{E}$ but with the charge acceleration $\mathrm{d}^2\mathbf{p}/\mathrm{d}t^2$ as a source term. This agrees with the intuition that the charge creates its own electromagnetic field in the environment when participating in electromagnetic interactions. To solve Eq. (\ref{eq:E_wave}), we first define a time-domain dyadic Green function for the dipole $\overline{\mathbf{G}}_\mathrm{dipole}$ as the solution to
\begin{equation}
\varepsilon(\mathbf{x})\varepsilon_0\frac{\partial^2 \overline{\mathbf{G}}_\mathrm{dipole}}{\partial t^2} + \frac{1}{\mu_0}\nabla\times\nabla\times \overline{\mathbf{G}}_\mathrm{dipole} = \delta(\mathbf{x})\overline{\mathbf{I}}\delta(t)
\end{equation}
together with outgoing boundary conditions, where $\overline{\mathbf{I}}$ is the three-dimensional identity matrix. The particular solution to Eq. (\ref{eq:E_wave}) can then be expressed as a convolution integral:
\begin{equation}
\mathbf{E}_\mathrm{dipole}(\mathbf{x},t) = -\int_{-\infty}^{+\infty} \overline{\mathbf{G}}_\mathrm{dipole}(\mathbf{x},t-t')\cdot\frac{\mathrm{d}^2\mathbf{p}(t')}{\mathrm{d}t'^2} \mathrm{d}t'
\end{equation}
This solution is denoted as $\mathbf{E}_\mathrm{dipole}$ because it can be interpreted as the field induced by the dipole.

We can now write the solution of Eq. (\ref{eq:E_wave}) as
\begin{equation}
\mathbf{E} = \mathbf{E}_\mathrm{dipole} + \mathbf{E}_\mathrm{ext}
\label{eq:E_decompose}
\end{equation}
Here $\mathbf{E}_\mathrm{ext}$ corresponds to the general solution in which the electromagnetic waves propagate freely as if the dipole were not there. Equation (\ref{eq:E_decompose}) thus describes the scattering process from the dipole, with $\mathbf{E}_\mathrm{dipole}$ corresponding to the scattered field. Inserting Eq. (\ref{eq:E_decompose}) as the driving field in Eq. (\ref{eq:LD_no_loss}), we obtain:
\begin{equation}
\frac{\mathrm{d}^2\mathbf{p}}{\mathrm{d}t^2} + \omega_0^2\mathbf{p} = \frac{q^2}{m} \mathbf{E}_{0,\mathrm{ext}} - \frac{q^2}{m}\int_{-\infty}^{+\infty} \overline{\mathbf{G}}_\mathrm{dipole}(\mathbf{x}=\mathbf{0},t-t')\cdot\frac{\mathrm{d}^2\mathbf{p}(t')}{\mathrm{d}t'^2} \mathrm{d}t'
\label{eq:LD_decompose}
\end{equation}
The second term on the right hand side describes the energy exchange between the dipole and the environment and provides radiation damping of the dipole.

The derivation remains exact up to this point. To arrive at an expression for $\kappa_\mathrm{rad}$, we further assume an isotropic response (i.e., $\overline{\mathbf{G}}_\mathrm{dipole} = G_\mathrm{dipole} \overline{\mathbf{I}}$) and switch to the frequency domain by $\partial/\partial t \rightarrow -i\omega$ with $\omega$ the angular frequency and $i^2 = -1$. Equation (\ref{eq:LD_decompose}) then becomes
\begin{equation}
\left[-\omega^2 + \omega_0^2 - \omega^2 \frac{q^2}{m}\hat{G}_\mathrm{dipole}(\mathbf{x}=\mathbf{0}, \omega)\right]\hat{\mathbf{p}} = \frac{q^2}{m}\hat{\mathbf{E}}_{0,\mathrm{ext}}
\end{equation}
where a hat denotes fields and quantities in the frequency domain.
If $\hat{\mathbf{G}}_\mathrm{dipole}$ varies slowly within a narrow frequency range near $\omega_0$, we can make the following approximation: 
\begin{equation}
-\omega^2\hat{G}_\mathrm{dipole} \approx -\omega_0^2\mathrm{Re}[\hat{G}_\mathrm{dipole}(\omega_0)] -i\omega \omega_0 \mathrm{Im}[\hat{G}_\mathrm{dipole}(\omega_0)]
\label{eq:markov}
\end{equation}
The real part of the Green function thus results in corrections to the resonance frequency that renormalize $\omega_0$. In practice, $\hat{G}_\mathrm{dipole}$ has a $1/|\mathbf{x}|$ divergence at $\mathbf{x}=\mathbf{0}$, and proper regularization procedures are required \cite{vries1998point} to incorporate the renormalization. On the other hand, $\mathrm{Im}[\hat{G}_\mathrm{dipole}(\omega_0)]$ remains finite. The approximation in Eq. (\ref{eq:markov}) is similar to applying the first Markov approximation to the system, where the density of states characterized by $\mathrm{Im}[\hat{G}_\mathrm{dipole}(\omega)]$ is replaced by a constant by setting $\omega = \omega_0$. By Fourier transforming back to the time domain, we arrive at the damped harmonic oscillator dynamics for the dipole:
\begin{equation}
\frac{\mathrm{d}^2\mathbf{p}}{\mathrm{d}t^2} + \left\{\frac{q^2}{m}\omega_0\mathrm{Im}[\hat{G}_\mathrm{dipole}(\mathbf{x}=\mathbf{0}, \omega_0)]\right\}\frac{\mathrm{d}\mathbf{p}}{\mathrm{d}t}  + \omega_0^2\mathbf{p} = \frac{q^2}{m} \mathbf{E}_{0,\mathrm{ext}}
\label{eq:dm_dynamic}
\end{equation}
We can then interpret the coefficient in the curly bracket as the radiation loss $\kappa_\mathrm{rad}$ near the renormalized resonance frequency $\omega_0$.

To compare the two formalisms with and without the self-field, we define two types of dipole polarizability in the frequency domain. The dynamic polarizability \cite{moskalensky2021point} is given by
\begin{equation}
\chi_\mathrm{dy}(\omega) = \frac{\hat{p}(\omega)}{\hat{E}_\mathrm{ext}(\omega)} = \frac{\omega_0^2}{\omega_0^2-\omega^2-i(\kappa_\mathrm{rad} + \kappa_\mathrm{nrad}) \omega} \frac{q^2}{m\omega_0^2}
\label{eq:chi_dynamic}
\end{equation}
This polarizability profile matches the form of Eq. (\ref{eq:LD_with_loss}).
The static polarizability \cite{moskalensky2021point} is given by
\begin{equation}
\chi_\mathrm{st}(\omega) = \frac{\hat{p}(\omega)}{\hat{E}_\mathrm{tot}(\omega)} = \frac{\omega_0^2}{\omega_0^2-\omega^2-i\kappa_\mathrm{nrad} \omega} \frac{q^2}{m\omega_0^2}
\label{eq:chi_static}
\end{equation}
which matches Eq. (\ref{eq:LD_no_loss}). The $\chi_\mathrm{st}(\omega)$ has a similar form to $\chi_\mathrm{dy}(\omega)$ but with the $\kappa_\mathrm{rad}$ term removed. The dynamic polarizability $\chi_\mathrm{dy}(\omega)$ depends on the environment and directly characterizes the total energy loss of the dipole. The static polarizability $\chi_\mathrm{st}(\omega)$, on the other hand, is an intrinsic property of the dipole and characterizes the non-radiative energy losses of the dipole independent of the environment. These two polarizabilities are related to the extinction and absorption powers of the dipole, which can be expressed as $W_\mathrm{ext} = \omega |\hat{E}_\mathrm{tot}|^2 \mathrm{Im}(\chi_\mathrm{dy})/2$ and $W_\mathrm{abs} = -\omega |\hat{p}|^2 \mathrm{Im}(\chi_\mathrm{st}^{-1})/2$, respectively \cite{moskalensky2021point}.

In summary, the two forms of the Lorentz-Drude model [Eqs. (\ref{eq:LD_no_loss}) and Eq. (\ref{eq:LD_with_loss})] differ in the way the driving field is interpreted. When using Eq. (\ref{eq:LD_no_loss}), the driving field is the total field, and there is no need to incorporate the radiative damping rate. On the other hand, when using Eq. (\ref{eq:LD_with_loss}), the driving field is the external field, and the radiative damping rate appears explicitly in the equation. This distinction is insignificant when $\kappa_\mathrm{rad} \ll \kappa_\mathrm{nrad}$, but becomes important in the radiative limit.

\subsection{Lorentz-Drude dipoles in a one-dimensional continuum} \label{ssec:H_example}

As an example of applying the formalism described in Section \ref{ssec:H_theory}, we consider Lorentz-Drude dipoles distributed uniformly in a plane and interact with light having an incident direction normal to the plane. This setup is a classical analog of a two-level atom coupled to a one-dimensional continuum \cite{shen2005coherent}. In the following, we demonstrate the perfect reflection of light on resonance in the radiative limit and connect the results with an input-output approach.

We assume that the dipoles are uniformly distributed in the $xy$ plane at $z=0$ with a surface density of $\sigma$ and are placed in vacuum ($\varepsilon(\mathbf{x}) = 1$). The wave equation of $\mathbf{E}$ can be found as
\begin{equation}
\varepsilon_0\frac{\partial^2 \mathbf{E}}{\partial t^2} + \frac{1}{\mu_0}\nabla\times\nabla\times \mathbf{E} = -\frac{\mathrm{d}^2\mathbf{p}}{\mathrm{d}t^2} \sigma\delta(z)
\end{equation}
which is similar to Eq. (\ref{eq:E_wave}) but with the point source replaced by a surface source.
Assuming that the light is linearly polarized in the $x$ direction and propagates in the $z$ direction, the wave equation can be further simplified as:
\begin{equation}
\varepsilon_0\frac{\partial^2 E_x}{\partial t^2} - \frac{1}{\mu_0}\frac{\partial^2 E_x}{\partial z^2} = -\frac{\mathrm{d}^2 p_x}{\mathrm{d}t^2} \sigma\delta(z)
\label{eq:wg_wave}
\end{equation}
where the subscript $x$ indicates the $x$ component. The dynamics for $p_x$ can be found from Eq. (\ref{eq:LD_no_loss}) as:
\begin{equation}
\frac{\mathrm{d}^2 p_x}{\mathrm{d}t^2} + \omega_0^2p_x = \frac{q^2}{m}E_x(z=0, t)
\label{eq:wg_dipole}
\end{equation}

Away from the plane $z = 0$, the fields can be expressed as superpositions of waves traveling in the $z$ direction. If all incident light comes from the $-z$ side, the general field solution can be found as
\begin{equation}
E_x(z,t) = \left\{
\begin{matrix}
E_\mathrm{in}(t-z/c) + E_\mathrm{r}(t+z/c), & z\leq0\\
E_\mathrm{t}(t-z/c), & z\leq0
\end{matrix}
\right.
\label{eq:wg_field_ansatz}
\end{equation}
where $E_\mathrm{in}$ is the incident field, $E_\mathrm{r}$ is the reflected field, $E_\mathrm{t}$ is the transmitted field, and $c = 1/\sqrt{\varepsilon_0\mu_0}$ is the speed of light in vacuum. By the continuity of the field, the field for $z \leq 0$ and $z \geq 0$ must match at $z = 0$ for all $t$. This implies that
\begin{equation}
E_\mathrm{t}(t-z/c) = E_\mathrm{in}(t-z/c) + E_\mathrm{r}(t-z/c)
\label{eq:wg_Et_sol}
\end{equation}
Integrating Eq. (\ref{eq:wg_wave}) across $z = 0$ and using Eq. (\ref{eq:wg_field_ansatz}), the field derivative difference across $z = 0$ can be found as
\begin{equation}
-\frac{2}{c}E_\mathrm{r}'=\sigma\frac{\mathrm{d}^2p_x}{\mathrm{d}t^2}
\label{eq:wg_field_jump}
\end{equation}
where a prime indicates derivative with respect to the argument. Using Eq. (\ref{eq:wg_field_jump}) to eliminate $E_\mathrm{r}$ from Eq. (\ref{eq:wg_dipole}) leads to
\begin{equation}
\frac{\mathrm{d}^2 p_x}{\mathrm{d}t^2} + \omega_0^2p_x = \frac{q^2}{m}\left(E_\mathrm{in} - \frac{\sigma c}{2}\frac{\mathrm{d} p_x}{\mathrm{d}t}\right)
\label{eq:wg_px}
\end{equation}
and its solution can be found as
\begin{equation}
\hat{p}_x(\omega) = \frac{\kappa_\mathrm{rad}}{\omega_0^2-\omega^2-i\kappa_\mathrm{rad}\omega}\frac{2\hat{E}_\mathrm{in}(\omega)}{\sigma c}
\label{eq:wg_px_freq}
\end{equation}
\begin{equation}
p_x(t) = \int_{-\infty}^t \frac{\kappa_\mathrm{rad}}{\omega_\mathrm{rad}} \exp\left[-\frac{\kappa_\mathrm{rad}}{2}(t-\tau)\right]\sin\left[\omega_\mathrm{rad}(t-\tau)\right]\frac{2E_\mathrm{in}(\tau)}{\sigma c}d\tau
\label{eq:wg_px_sol}
\end{equation}
where we have defined the radiation rate and renormalized resonant frequency for the system as
\begin{equation}
\kappa_\mathrm{rad} = \frac{q^2\sigma c}{2m}
\end{equation}
\begin{equation}
\omega_\mathrm{rad} = \sqrt{\omega_0^2-\frac{\kappa_\mathrm{rad}^2}{4}}
\end{equation}
Using Eq. (\ref{eq:wg_px_freq}), the reflection and transmission coefficient can be found as
\begin{equation}
\frac{\hat{E}_\mathrm{r}(\omega)}{\hat{E}_\mathrm{in}(\omega)} = \frac{i\kappa_\mathrm{rad}\omega}{\omega_0^2-\omega^2-i\kappa_\mathrm{rad}\omega}
\label{eq:wg_coeff_R}
\end{equation}
\begin{equation}
\frac{\hat{E}_\mathrm{t}(\omega)}{\hat{E}_\mathrm{in}(\omega)} = \frac{\omega_0^2-\omega^2}{\omega_0^2-\omega^2-i\kappa_\mathrm{rad}\omega}
\label{eq:wg_coeff_T}
\end{equation}

If the input light oscillates at the resonance frequency of the dipole, i.e. $E_\mathrm{in} = E_\mathrm{in}(t=0)\cos(\omega_0 t)$, then Eq. (\ref{eq:wg_coeff_R}) indicates that $E_\mathrm{r} = -E_\mathrm{in}(t=0)\cos(\omega_0 t)$ and Eq. (\ref{eq:wg_coeff_T}) indicates that $E_\mathrm{t} = 0$. Thus, perfect reflection can be achieved on resonance when the dipole does not contain intrinsic losses. Here, perfect reflection occurs at the dipole resonance frequency $\omega_0$ instead of the renormalized resonance frequency $\omega_\mathrm{rad}$. We note that no additional assumptions have been introduced in this derivation. Specifically, we have not used any rotating wave approximations or assumed $\kappa_\mathrm{rad}$ to be small.

Finally, we connect our derivations to an input-output approach \cite{gardiner1985input, gardiner2000quantum, fan2010input-output}, which also illustrates the role of the total field in the system. We use both the weak-coupling approximation ($\kappa_\mathrm{rad} \ll \omega_0$ and $\omega_\mathrm{rad} \approx \omega_0$) and the rotating-wave approximation (only frequency components around the resonance will be considered), as is commonly assumed in the input-output approach. We write $p_x = \mathrm{Re}[p_x^{(+)}]$ where $(+)$ indicates complex quantities that contain only positive frequency components. Eq. (\ref{eq:wg_px}) can then be simplified to
\begin{equation}
\frac{\mathrm{d} p_x^{(+)}}{\mathrm{d}t} = \left(-\frac{\kappa_\mathrm{rad}}{2}- i\omega_0\right) p_x^{(+)} + \frac{i\kappa_\mathrm{rad}}{\omega_\mathrm{rad}}\frac{E_\mathrm{in}^{(+)}}{\sigma c}
\label{eq:wg_io_in}
\end{equation}
This clearly shows the presence of the radiative loss term of the dipole when the driving field is $E_\mathrm{in}$, i.e. the external field. However, by considering a time-reversed process, we can also take $E_\mathrm{r}$ and $E_\mathrm{t}$ fields as time-reversed inputs to $p_x$, and write
\begin{equation}
\frac{\mathrm{d} p_x^{(+)}}{\mathrm{d}t} = \left(\frac{\kappa_\mathrm{rad}}{2}- i\omega_0\right) p_x^{(+)} + \frac{i\kappa_\mathrm{rad}}{\omega_\mathrm{rad}}\frac{E_\mathrm{r}^{(+)}+E_\mathrm{t}^{(+)}}{\sigma c}
\label{eq:wg_io_out}
\end{equation}
Equations (\ref{eq:wg_io_in}) and (\ref{eq:wg_io_out}) can be combined to produce
\begin{equation}
\frac{\mathrm{d} p_x^{(+)}}{\mathrm{d}t} = - i\omega_0 p_x^{(+)} + \frac{i\kappa_\mathrm{rad}}{2\omega_\mathrm{rad}}\frac{E_\mathrm{in}^{(+)}+E_\mathrm{r}^{(+)}+E_\mathrm{t}^{(+)}}{\sigma c}
\label{eq:wg_io_sym}
\end{equation}
where the driving term is now symmetrized with respect to inputs and outputs. From Eq. (\ref{eq:wg_field_ansatz}), the average of the input and output fields is exactly the total field at the dipole. Therefore, no loss terms appear explicitly in the input-output formalism if the driving term is replaced by the total field. The discussion here shows the consistency of our approach with the standard input-output approach.

\section{Implementation of Lorentz-Drude dipoles} \label{sec:implement}

In the previous section, we showed that a dipole at the radiative limit can be modeled using Eq. (\ref{eq:LD_no_loss}) where the driving field is taken to be the total field at the dipole. In addition to the theoretical interests, as illustrated in the previous section, the use of Eq. (\ref{eq:LD_no_loss}) is also important in numerical simulations of a radiative dipole embedded in a photonic structure. Common numerical methods for solving Maxwell's equations, such as finite-difference time-domain (FDTD) methods \cite{taflove2010computational}, directly produce the total field at the dipole. In this section, we show that incorporating Eq. (\ref{eq:LD_no_loss}) into the numerical solver allows us to treat the modification of the radiative dynamics of the dipole due to the presence of the photonic structures, providing a direct time-domain approach for modeling the Purcell enhancement effect.
We first analytically calculate the Green function on the Yee grid in Section \ref{ssec:imp_green}, which is necessary to describe the field propagation on the discretized grid. After that, we provide two possible approximations of point dipoles on the Yee grid in Sections \ref{ssec:imp_anisotropic} and \ref{ssec:imp_isotropic}, and relate their physical properties (radiation frequency and rate) to the input parameters using the Green function.

\subsection{Yee grid and the frequency-domain Green function}
\label{ssec:imp_green}

In standard FDTD simulations, the Yee grids discretize the space into cubic voxels, where the electric ($\mathbf{E}$) field components are sampled at the midpoint of the cube edge (the ``$E$ points''), and the magnetic ($\mathbf{H}$) field components are sampled at the center of the cube face (the ``$H$ points''). If the coordinate unit is chosen as the grid length in the respective direction, then each grid point (cube corner) has a coordinate in the form of $(u, v, w)$ where $u$, $v$ and $w$ are all integers. The $E$ fields are sampled at the points where one coordinate component is a half-integer and the other two are integers, and the $H$ fields are sampled at the points where two coordinate components are half-integers and the other one is an integer. For simplicity, we assume that the grid length in each dimension is identical, denoted as $a$.

We first consider the radiation of a dipole source in free space. We note that the dipole source within this subsection refers to a source where the oscillation of the dipole is externally imposed, which is different from the Lorentz-Drude dipole discussed throughout the paper that has internal dynamics as modeled, for example, by Eq. (\ref{eq:LD_no_loss}). To discuss the dynamics of a dipole source on the Yee grid in the FDTD method, it is useful first to consider the frequency domain wave equation, which can be written as
\begin{equation}
\nabla\times\nabla\times\mathbf{E} - \frac{\omega^2}{c^2} \mathbf{E} = i\omega\mu_0 \mathbf{J}
\end{equation}
where $\mathbf{E}$ is the electric field, $\nabla\times$ is the curl operator, $\omega$ is the angular frequency, $c$ is the speed of light in vacuum, $\mu_0$ is the vacuum permeability and $\mathbf{J}$ is the current density. In FDTD, the curl is replaced by its discretized version $\nabla_a\times$, where
\begin{align}
& (\nabla_a\times \mathbf{E})_z(u,v,w) \nonumber\\
&=\frac{1}{a}\left[E_y\left(u+\frac{1}{2}, v, w\right) - E_y\left(u-\frac{1}{2}, v, w\right) - E_x\left(u, v+\frac{1}{2}, w\right) + E_x\left(u, v-\frac{1}{2}, w\right)\right]
\end{align}
and the $x$ and $y$ components can be obtained by interchanging the directions cyclically. As such, the frequency-domain Green function on the Yee grid reads
\begin{equation}
a^2\nabla_a\times\nabla_a\times\mathbf{G}_\mathrm{Yee} - (\alpha^2 + i0^+) \mathbf{G}_\mathrm{Yee} = \delta\left(0,0,\frac{1}{2}\right)\mathbf{e}_z
\label{eq:FDTD_GYee_def}
\end{equation}
where the Kronecker delta is $1$ on the point $(0,0,1/2)$ and $0$ elsewhere, $\mathbf{e}_z$ is the unit vector along $z$, $\alpha = \omega a/c$, $0^+$ is an infinitesimal positive number that selects the outgoing wave, and the equation has been scaled by $a^2$ so that $\mathbf{G}_\mathrm{Yee}$ is dimensionless. We note that $2\pi/\alpha$ gives the number of grid points in a single wavelength. 

The Green function can be solved in the wavevector domain \cite{ma2005discrete, kastner2006multidimensional}. The Fourier transform of $\mathbf{G}_\mathrm{Yee}$ reads
\begin{equation}
\tilde{\mathbf{G}}_\mathrm{Yee}(\mathbf{k}) = \sum_{u,v,w} \mathbf{G}_\mathrm{Yee}(u,v,w) \exp\left[-i k_x u - i k_y v - i k_z \left(w-\frac{1}{2}\right)\right]
\end{equation}
where $\mathbf{k} = (k_x, k_y, k_z)$ is the grid wavevector. The equation for $\tilde{\mathbf{G}}_\mathrm{Yee}(\mathbf{k})$ can be found as
\begin{equation}
\begin{bmatrix}
s_y^2+s_z^2 & -s_x s_y & -s_x s_z \\
-s_x s_y & s_x^2+s_z^2 & -s_y s_z \\
-s_x s_z & -s_y s_z & s_x^2+s_y^2
\end{bmatrix}
\tilde{\mathbf{G}}_\mathrm{Yee}(\mathbf{k})
- (\alpha^2 + i0^+) \tilde{\mathbf{G}}_\mathrm{Yee}(\mathbf{k})
= \begin{bmatrix}
0 \\ 0 \\ 1
\end{bmatrix}
\end{equation}
with $s_x = 2\sin(k_x/2)$, $s_y = 2\sin(k_y/2)$ and $s_y = 2\sin(k_z/2)$. Its solution is
\begin{equation}
\tilde{\mathbf{G}}(\mathbf{k}) = \frac{1}{\alpha^2[\alpha^2+i0^+-(s_x^2+s_y^2+s_z^2)]}
\begin{bmatrix}
s_x s_z \\ s_y s_z \\ s_z^2-\alpha^2
\end{bmatrix}
\end{equation}
The Green function can then be recovered using the inverse Fourier transform:
\begin{equation}
\mathbf{G}_\mathrm{Yee}(u,v,w) = \frac{1}{8\pi^3}\int_{-\pi}^{\pi} \mathrm{d}k_x \int_{-\pi}^{\pi} \mathrm{d}k_y \int_{-\pi}^{\pi} \mathrm{d}k_z \tilde{\mathbf{G}}_\mathrm{Yee}(\mathbf{k}) \exp\left[i k_x u + i k_y v + i k_z \left(w-\frac{1}{2}\right)\right]
\end{equation}

To calculate $\mathbf{G}_\mathrm{Yee}$, we switch to a spherical coordinate system by defining
\begin{align}
s_x = 2\sin(k_x/2) &= \rho \sin\mu \cos\nu \\
s_y = 2\sin(k_y/2) &= \rho \sin\mu \sin\nu \\
s_z = 2\sin(k_z/2) &= \rho \cos\mu
\end{align}
with $0 \leq \mu \leq \pi$ and $0 \leq \nu \leq 2\pi$. The full expression for $\mathbf{G}_\mathrm{Yee}$ reads
\begin{align}
\mathbf{G}_\mathrm{Yee}(u,v,w) &= -\frac{1}{8\pi^3\alpha^2}\iiint \mathrm{d}\nu \mathrm{d}\mu \mathrm{d}\rho \nonumber\\
&\times \frac{8\rho^2\sin\mu}{\sqrt{4-\rho^2\cos^2\mu}\sqrt{4-\rho^2\sin^2\mu\cos^2\nu}\sqrt{4-\rho^2\sin^2\mu\sin^2\nu}} \nonumber\\
&\times \frac{\exp[i k_x u + i k_y v + i k_z (w-1/2)]}{\rho^2-\alpha^2-i0^+}
\begin{bmatrix}
\rho^2 \cos\mu \sin\mu \cos\nu \\ \rho^2 \cos\mu \sin\mu \sin\nu \\ \rho^2 \cos^2\mu -\alpha^2
\end{bmatrix}
\end{align}
The integration domain is the cube $|\rho \cos\mu \cos\nu| \leq 2$, $|\rho \cos\mu \sin\nu| \leq 2$ and $|\rho \sin\mu| \leq 2$.

To explicitly incorporate the outgoing wave boundary condition and remove the $0^+$ from the expressions, we use the fact that
\begin{equation}
\frac{1}{\rho-i0^+} = i\pi\delta(\rho) + \mathrm{PV}\frac{1}{\rho}
\end{equation}
where PV is the principal value for the integration. For $\mathbf{G}_\mathrm{Yee}$, the pole occurs at $\rho=\alpha$, representing the traveling waves on the Yee grid. We now assume that $\alpha < 2$ (i.e. more than $\pi$ grid points per wavelength), a reasonable constraint for practical FDTD simulations. With this assumption, the real and imaginary parts of $\mathbf{G}_\mathrm{Yee}$ can be obtained as
\begin{align}
\mathrm{Re}\mathbf{G}_\mathrm{Yee}(u,v,w) &= -\frac{1}{8\pi^3\alpha^2}\int_{0}^{2\pi} \mathrm{d}\nu \int_{0}^{\pi} \mathrm{d}\mu \times \mathrm{PV}\int_0^{2/\mathrm{max}(|\sin\mu \cos\nu|,|\sin\mu \sin\nu|,|\cos\mu|)} \mathrm{d}\rho \nonumber\\
&\times \frac{8\rho^2\sin\mu}{\sqrt{4-\rho^2\sin^2\mu\cos^2\nu}\sqrt{4-\rho^2\sin^2\mu\sin^2\nu}\sqrt{4-\rho^2\cos^2\mu}} \nonumber\\
&\times \frac{\cos(k_x u)\cos(k_y v)\cos[k_z (w-1/2)]}{\rho^2-\alpha^2}
\begin{bmatrix}
\rho^2 \cos\mu \sin\mu \cos\nu \\ \rho^2 \cos\mu \sin\mu \sin\nu \\ \rho^2 \cos^2\mu -\alpha^2
\end{bmatrix}
\end{align}
\begin{align}
\mathrm{Im}\mathbf{G}_\mathrm{Yee}(u,v,w) &= \frac{\alpha}{2\pi^2}\int_{0}^{2\pi} \mathrm{d}\nu \int_{0}^{\pi} \mathrm{d}\mu \nonumber\\
&\times \frac{\sin^2\mu}{\sqrt{4-\alpha^2\sin^2\mu\cos^2\nu}\sqrt{4-\alpha^2\sin^2\mu\sin^2\nu}\sqrt{4-\alpha^2\cos^2\mu}} \nonumber\\
&\times \cos(k_x u)\cos(k_y v)\cos[k_z (w-1/2)]|_{\rho=\alpha}
\begin{bmatrix}
-\cos\mu \cos\nu \\ -\cos\mu \sin\nu \\ \sin\mu
\end{bmatrix}
\end{align}
where we have used mirror symmetry to replace the exponential factors with cosines.

In practice, FDTD simulations require a large number of grids per wavelength to maintain accuracy, and $\alpha \ll 1$. As such, we can approximate $\mathbf{G}_\mathrm{Yee}(u,v,w)$ as a series in $\alpha$ to avoid performing integrations for each $\alpha$. For the $z$-polarized dipole source located at $(0,0, 1/2)$ as indicated in Eq. (\ref{eq:FDTD_GYee_def}), some values of $\mathbf{G}_\mathrm{Yee}$ located near the source point are listed below.
\begin{align}
G_z\left(0,0,\frac{1}{2}\right) &= -\frac{1}{3}\alpha^{-2} + 0.168487 + \frac{i}{6\pi} \alpha + O(\alpha^2) \label{eq:GYee_value_1}\\
G_z\left(0,0,-\frac{1}{2}\right) &= 0.123492\alpha^{-2} +0.084243 + \frac{i}{6\pi} \alpha + O(\alpha^2) \label{eq:GYee_value_2}\\
G_x\left(\frac{1}{2},0,0\right) &= -0.135794\alpha^{-2} -0.021061 + O(\alpha^2) \label{eq:GYee_value_3}
\end{align}

\subsection{The anisotropic dipole}
\label{ssec:imp_anisotropic}

We now discuss the modeling of the dipole dynamics and Purcell effects by incorporating Eq. (\ref{eq:LD_no_loss}) into the FDTD simulations. In these simulations, we typically assume a Lorentz-Drude dipole, which has a renormalized resonant frequency $\omega_\mathrm{rad}$ and a radiative decay rate $\gamma_\mathrm{rad}$ when placed in vacuum. We then consider how the resonant frequency and the radiative rate are modified when such a dipole is placed in a photonic structure. The renormalized resonant frequency $\omega_\mathrm{rad}$ differs from the bare resonant frequency $\omega_0$ of the Lorentz-Drude dipole even when surrounded by vacuum. However, when specifying the parameters of Eq. (\ref{eq:LD_no_loss}), the information of $\omega_0$ is required. Therefore, in this and the next section, we provide a discussion on how to obtain $\omega_0$ and the other parameters of Eq. (\ref{eq:LD_no_loss}) by analytically considering the behavior of a Lorentz-Drude dipole placed in a Yee lattice.

Since the distribution of dielectrics is represented in FDTD simulations by assigning the permittivity to the $E$ points on the grid, the closest approximation of a point dipole would be to assign the Lorentz-Drude profile to a single $E$ point. Because the resulting dipole can only be polarized in a single direction aligned with the Yee grid, we term this implementation of the point dipole as the anisotropic dipole. This method has been briefly discussed in a phenomenological way in \cite{schelew2017self-consistent}. Here we provide an analytical description of the radiation induced by the anisotropic dipole in order to determine the Lorentz-Drude parameters used in simulations.

We assume that the anisotropic dipole is located at $(0,0,1/2)$ with a relative permittivity of $\varepsilon(\omega)$ and is locally surrounded by vacuum. The frequency-domain wave equation including the anisotropic dipole reads
\begin{equation}
\nabla_a\times\nabla_a\times\mathbf{E} - \frac{\omega^2}{c^2} \left[1 + (\varepsilon - 1)\delta\left(0,0,\frac{1}{2}\right)\right]\mathbf{E} = 0
\end{equation}
The field can be separated into two contributions $\mathbf{E} = \mathbf{E}_\mathrm{ext} + \mathbf{E}_\mathrm{dipole}$, where $\mathbf{E}_\mathrm{ext}$ describes the external field and $\mathbf{E}_\mathrm{dipole}$ is the field scattered from the anisotropic dipole. These are required to satisfy the following:
\begin{equation}
\nabla_a\times\nabla_a\times\mathbf{E}_\mathrm{ext} - \frac{\omega^2}{c^2}\mathbf{E}_\mathrm{ext} = 0
\end{equation}
\begin{equation}
\nabla_a\times\nabla_a\times\mathbf{E}_\mathrm{dipole} - \frac{\omega^2}{c^2}\mathbf{E}_\mathrm{dipole} = \frac{\omega^2}{c^2}(\varepsilon-1)\left(\mathbf{E}_\mathrm{ext} + \mathbf{E}_\mathrm{dipole}\right)\delta\left(0,0,\frac{1}{2}\right)
\end{equation}
As such, the scattered field can be solved in terms of the Green's function $\mathbf{G}_\mathrm{Yee}$ as:
\begin{equation}
\mathbf{E}_\mathrm{dipole} = \left[G_0 - \frac{1}{\alpha^2}\frac{1}{\varepsilon - 1}\right] ^{-1}E_{z,\mathrm{ext}}\left(0,0,\frac{1}{2}\right)\mathbf{G}_\mathrm{Yee}
\end{equation}
where $G_0 \equiv G_z(0,0,1/2)$ is the value of $\mathbf{G}_\mathrm{Yee}$ on the source point. Resonance will occur if the complex frequency satisfies
\begin{equation}
(\varepsilon - 1)\alpha^2 G_0 = 1
\label{eq:reson_anisotropic}
\end{equation}
This holds for general $\varepsilon(\omega)$ profiles. We note that $\alpha$ and $G_0$ also implicitly depend on $\omega$.

For a dipole at the radiative limit with a Lorentz-Drude profile, the relative permittivity is given by
\begin{equation}
\varepsilon = 1 + \Delta \varepsilon \frac{\omega_0^2}{\omega_0^2-\omega^2}
\label{eq:LD_dielectric}
\end{equation}
with $\Delta \varepsilon$ the static susceptibility at $\omega = 0$ and $\omega_0$ the bare resonant frequency for the Lorentz-Drude profile. In time domain, this is realized by a harmonic oscillator model:
\begin{equation}
\frac{\mathrm{d}^2P}{\mathrm{d}t^2} + \omega_0^2 P = \Delta \varepsilon \omega_0^2 \varepsilon_0 E
\end{equation}
where $\varepsilon_0$ is the vacuum permittivity and $P$ is the polarization field [compare with Eq. (\ref{eq:LD_no_loss})]. We note that the driving electric field $E$ represents the total field strength at the location of the dipole. As such, radiation loss is not included in Eq. (\ref{eq:LD_dielectric}), consistent with the Lorentz-Drude profile being an intrinsic material property.

Applying the resonance condition Eq. (\ref{eq:reson_anisotropic}) to the Lorentz-Drude profile Eq. (\ref{eq:LD_dielectric}) leads to a self-consistent equation for the complex resonance frequency $\omega_\mathrm{res}$:
\begin{equation}
\omega_\mathrm{res} = \omega_0\sqrt{1+\Delta\varepsilon\left(\frac{1}{3} - 0.168487\alpha_\mathrm{res}^2 - \frac{i}{6\pi} \alpha_\mathrm{res}^3 + O(\alpha_\mathrm{res}^4)\right)}
\end{equation}
with $\alpha_\mathrm{res} = \omega_\mathrm{res}a/c$. From here, the dipole radiation frequency and the dipole radiation rate can be found as $\omega_\mathrm{rad} \equiv \mathrm{Re}\ \omega_\mathrm{res}$ and $\kappa_\mathrm{rad} \equiv -2\ \mathrm{Im}\ \omega_\mathrm{res}$, respectively. Under the additional assumption that $\kappa_\mathrm{rad} \ll \omega_\mathrm{rad}$, we can approximate $\alpha_\mathrm{res} \approx \omega_\mathrm{rad}a/c$ and arrive at
\begin{equation}
\kappa_\mathrm{rad} \approx \frac{a^3}{6\pi c^3}\frac{\Delta\varepsilon}{1 + \Delta\varepsilon/3 -0.168487\Delta\varepsilon\omega_\mathrm{rad}^2a^2/c^2}\omega_\mathrm{rad}^4
\label{eq:kappa_anisotropic}
\end{equation}
\begin{equation}
\omega_0 \approx \omega_\mathrm{rad}\left[1+\Delta\varepsilon\left(\frac{1}{3} - 0.168487\frac{\omega_\mathrm{rad}^2a^2}{c^2} \right)\right]^{-1/2}
\label{eq:omega_anisotropic}
\end{equation}

We provide a physical interpretation of Eq. (\ref{eq:kappa_anisotropic}) by comparing it to the radiation rate of a point dipole in vacuum,
\begin{equation}
\kappa_\mathrm{rad} = \frac{\chi\omega_\mathrm{rad}^4}{6\pi \varepsilon_0 c^3}
\end{equation}
with $\chi$ the dipole polarizability. The two rates agree with each other to lowest orders of $\alpha$ if the dipole is taken as a sphere with volume $a^3$ and constant susceptibility $\Delta\epsilon$, such that $\chi = \varepsilon_0 a^3 \Delta\epsilon / (1 + \Delta\epsilon/3)$. The effective volume of the dipole is equal to the volume of a voxel, which can be generalized to Yee grids with unequal grid lengths in each direction. However, the effective spherical shape is due to the octahedral symmetry of the Yee grid with identical grid lengths in each dimension, rather than the dipole itself (which can only be polarized in the $z$ direction).

We now discuss the numerical procedure for implementing a dipole in FDTD based on Eqs. (\ref{eq:kappa_anisotropic}) and (\ref{eq:omega_anisotropic}). We assume that a point dipole is characterized by its radiating frequency $\omega_\mathrm{rad}$ and rate $\kappa_\mathrm{rad}$ in vacuum. To simulate this dipole in FDTD, the parameters in the Lorentz-Drude profile Eq. (\ref{eq:LD_dielectric}) should be determined. From Eq. (\ref{eq:kappa_anisotropic}), $\Delta\varepsilon$ can be solved as
\begin{equation}
\Delta\varepsilon = \frac{6\pi c^3}{a^3\omega_\mathrm{rad}^4-2\pi c^3\kappa_\mathrm{rad}}\kappa_\mathrm{rad}
\end{equation}
where higher-order terms of $\alpha$ has been neglected. Then the Lorentz-Drude oscillation frequency $\omega_0$ can be found using Eq. (\ref{eq:omega_anisotropic}). Here $\omega_0$ is different from $\omega_\mathrm{rad}$ due to the frequency renormalization induced by the self-field quantified by $G_0$. While the simulation parameters have been determined from dipole characteristics in vacuum, they are intrinsic material parameters, and are thus applicable even if the same dipole is surrounded by dielectric. This allows comparing the radiation properties of dipoles in different electromagnetic environments.

\subsection{The isotropic dipole}
\label{ssec:imp_isotropic}

The anisotropic dipole has the disadvantage that it can only be polarized in one direction aligned with the Yee grid. To simulate a dipole that responds to electric fields in all directions, we introduce the isotropic dipole setup, where all six $E$ points surrounding a single grid point are assigned a Lorentz-Drude profile. The six oscillators in the isotropic dipole support six oscillation modes, which can be classified by their symmetries as a monopole mode, three dipole modes, and two quadrupole modes. The monopole mode cannot be excited if there are no overlapping sources due to the divergence-free condition of the $E$ field, and the quadrupole modes are greatly suppressed by the continuity of the $E$ field if $\alpha$ is small enough. As such, the three dipole modes (one aligned with each grid direction) and their radiations are the main components of the scattered field from the isotropic dipole.

In addition to the assumption of equal grid length ($a$) in each dimension, we also assume that the same permittivity function $\epsilon$ is assigned to the six points $(\pm 1/2,0,0)$, $(0,\pm 1/2,0)$, and $(0,0,\pm 1/2)$. Using the same method from Section \ref{ssec:imp_anisotropic}, the resonance condition for the isotropic dipole can be found as
\begin{equation}
(\varepsilon - 1)\alpha^2 \left[G_z\left(0,0,\frac{1}{2}\right) + G_z\left(0,0,-\frac{1}{2}\right) \right] = 1
\end{equation}
where the source of the Green function is located at $(0,0,1/2)$ and the $G_z$ values have been calculated in Eqs. (\ref{eq:GYee_value_1}) and (\ref{eq:GYee_value_2}).

For the Lorentz-Drude profile Eq. (\ref{eq:LD_dielectric}), the relation between the simulation parameters $\omega_0$ and $\Delta\varepsilon$ and the observable dipole radiation characteristics $\omega_\mathrm{rad}$ and $\kappa_\mathrm{rad}$ can be summarized as
\begin{equation}
\kappa_\mathrm{rad} \approx \frac{a^3}{3\pi c^3}\frac{\Delta\varepsilon}{1 + 0.209842\Delta\varepsilon -0.252731\Delta\varepsilon\omega_\mathrm{rad}^2a^2/c^2}\omega_\mathrm{rad}^4
\label{eq:kappa_isotropic}
\end{equation}
\begin{equation}
\omega_0 \approx \omega_\mathrm{rad}\left[1+\Delta\varepsilon\left(0.209842 - 0.252731\frac{\omega_\mathrm{rad}^2a^2}{c^2} \right)\right]^{-1/2}
\label{eq:omega_isotropic}
\end{equation}
Compared to the anisotropic dipole, the effective volume for the dipole has been doubled ($2a^3$) but the effective geometry can no longer be considered as a sphere.

\section{Simulation results} \label{sec:examples}

In this section, we validate our analytical results from Sections \ref{sec:Hamiltonian} and \ref{sec:implement} by performing FDTD simulations in different scenarios with both anisotropic and isotropic dipole implementations. All FDTD simulations are performed using the Tidy3D solver from Flexcompute \cite{tidy3d}.

\subsection{Dipoles in free space}

To realize a point dipole with a given radiation frequency $\omega_\mathrm{rad}$ and a radiation rate $\kappa_{rad}$, Eqs. (\ref{eq:kappa_anisotropic}-\ref{eq:omega_anisotropic}) provides simulation parameters $\omega_0$ and $\Delta\varepsilon$ for the anisotropic dipole [Eqs. (\ref{eq:kappa_isotropic}-\ref{eq:omega_isotropic}) for the isotropic dipole]. In the following, we verify these relations by simulating the dipoles radiating in free space.

\begin{figure}
\centering
\includegraphics[width=132mm]{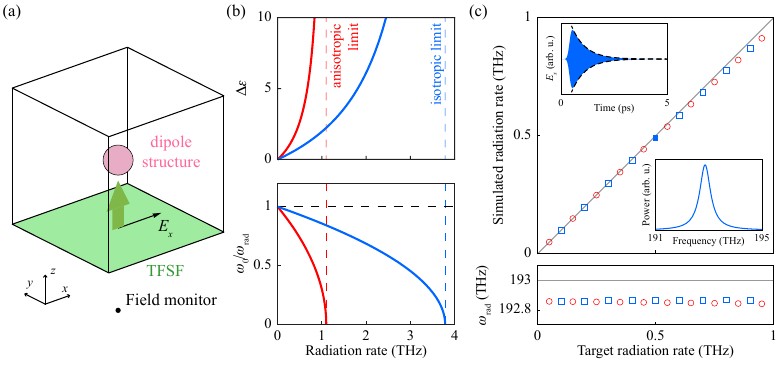}
\caption{Simulation of dipoles in free space.
(a) Simulation setup (not to scale). A complete geometric description of the simulation domain can be found in the main text.
(b) $\Delta\varepsilon$ (top panel) and $\omega_0$ (bottom panel) of anisotropic (red) and isotropic (blue) dipoles as a function of the radiation rate $\kappa_\mathrm{rad}$. The maximum possible $\kappa_\mathrm{rad}$ for each dipole implementation are marked with vertical dashed lines. Other parameters are $\omega_\mathrm{rad} = 2\pi \times 193$ THz and $a = 0.08$ $\mu$m.
(c) Comparison of simulated $\kappa_\mathrm{rad}$ (top panel) and $\omega_\mathrm{rad}$ (bottom panel) to target values (gray lines). Red circles are for anisotropic dipoles and blue squares are for isotropic dipoles.
Inset shows the time-dependent $E_x$ field at the monitor and the radiation spectrum for an isotropic dipole with $\kappa_\mathrm{rad} = 2\pi \times 0.5$ THz.
The field decay rate agrees with the spectral width.
}
\label{fig:sim_free}
\end{figure}

We construct a dipole radiating in free space at $\omega_\mathrm{rad} = 2\pi \times 193$ THz (wavelength $\lambda \approx 1553$ nm) with varying radiation rates [Fig. \ref{fig:sim_free}(a)]. The simulation domain is $8\ \mu\mathrm{m} \times 8\ \mu\mathrm{m} \times 8\ \mu\mathrm{m}$ and is padded with perfectly matched layers to absorb outgoing radiation. We fix $a = 0.08$ $\mu$m which is $\sim \lambda/20$ ($\alpha \approx 0.32$). The normalized Courant factor \cite{taflove2010computational} is chosen as $0.5$ to improve simulation accuracy. For the anisotropic dipole, a spherical structure with diameter $a/2$ is placed at the point $(a/2,0,0)$ so that the Lorentz-Drude profile can be assigned to the same point on the Yee grid. For the isotropic dipole, the spherical structure instead has a diameter of $3a/2$ and is centered at $(0,0,0)$. The quantities $\omega_0$ and $\Delta\epsilon$ are solved from $\omega_\mathrm{rad}$ and $\kappa_\mathrm{rad}$ using Eqs. (\ref{eq:kappa_anisotropic}-\ref{eq:omega_anisotropic}) for the anisotropic dipole or Eqs. (\ref{eq:kappa_isotropic}-\ref{eq:omega_isotropic}) for the anisotropic dipole and then supplied to the simulation. To excite the dipole, we launch a Gaussian pulse with $x$ polarization from a total-field-scattered-field (TFSF) source that encloses the dipole. A field monitor is placed $0.8\ \mu\textrm{m}$ away in the perpendicular direction of the polarization to record the time-dependent field and extract the radiation rate.

The theoretical dependence of $\omega_0$ and $\Delta\varepsilon$ on $\kappa_\mathrm{rad}$ is shown in Fig. \ref{fig:sim_free}(b). The curves shown in the plot allow finding the simulation parameters from $\kappa_\mathrm{rad}$. The strong self-field leads to $\omega_0$ being smaller than $\omega_\mathrm{rad}$. There is also an upper limit for $\kappa_\mathrm{rad}$ due to the small effective volume of the dipole (a single voxel for anisotropic dipoles and two voxels for isotropic dipoles).

The simulated $\kappa_\mathrm{rad}$ and $\omega_\mathrm{rad}$ can be found in Fig. \ref{fig:sim_free}(c). For higher radiation rates, $\kappa_\mathrm{rad}$ and $\omega_\mathrm{rad}$ can be found from the spectral linewidth and center after a Fourier transform. For lower radiation rates, the field does not decay fast enough within the simulation time window; instead, $\kappa_\mathrm{rad}$ and $\omega_\mathrm{rad}$ are obtained directly from the exponentially decaying part of the field in the time domain. The two methods agree at intermediate radiation rates [Insets of Fig. \ref{fig:sim_free}(c)]. Although $\omega_0$ and $\Delta\varepsilon$ change significantly with respect to $\kappa_\mathrm{rad}$, the simulated $\kappa_\mathrm{rad}$ shows excellent agreement with the target values. Deviations for larger $\kappa_\mathrm{rad}$ can be attributed to truncations of $G_0$ with respect to $\alpha$. The simulated $\omega_\mathrm{rad}$ have a small constant offset from the target value. This is attributed to the discretized time step of the FDTD method, as numerical evidence suggests that the error grows quadratically with respect to the Courant factor for fixed $a$. We note that while it is possible to include this effect in the Green function analysis, the necessary modifications depend on the detailed implementation of the dynamics for the polarization field.

\subsection{Two-dimensional dipole arrays}

A notable effect for dipoles without intrinsic loss is that they completely reflect incoming on-resonance waves when coupled to a one-dimensional waveguide \cite{shen2005coherent}, as discussed in Sec. \ref{ssec:H_example}. This generalizes to a two-dimensional dipole array and normal incident wave if only zeroth-order diffraction occurs \cite{shahmoon2017cooperative}. The two-dimensional array can be simulated by placing a single dipole in the simulation domain and replacing two pairs of opposite boundaries with periodic boundary conditions. This example demonstrates the difference between radiation and intrinsic loss, since any intrinsic loss for the dipole would remove electromagnetic energy from the system, leading to a reflection coefficient smaller than unity. In addition, as the dipole is physically excited by the sources, the implementations here allow direct extraction of the transmission and reflection spectra.

\begin{figure}
\centering
\includegraphics[width=132mm]{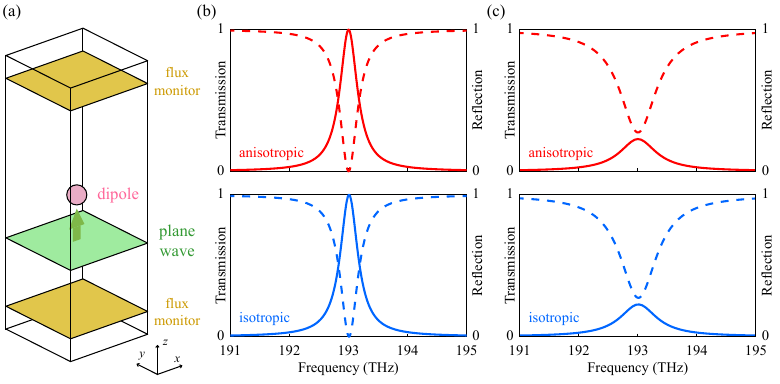}
\caption{Simulation of dipole arrays.
(a) Simulation setup (not to scale). A complete geometric description of the simulation domain can be found in the main text.
(b) Transmission (dashed curves) and reflection (solid curves) coefficients for lossless dipole arrays. Top and bottom panels use the anisotropic and isotropic dipole implementations, respectively.
(c) Transmission (dashed curves) and reflection (solid curves) coefficients for lossy dipole arrays, where the intrinsic loss is equal to the radiation loss in free space. Top and bottom panels use the anisotropic and isotropic dipole implementations, respectively.
}
\label{fig:sim_array}
\end{figure}

For this simulation, the $x$ and $y$ boundaries have periodic boundary conditions, and the simulation domain is $0.8\ \mu\mathrm{m} \times 0.8\ \mu\mathrm{m} \times 8\ \mu\mathrm{m}$ which also determines the dipole spacing in the array. The dipole has $a = 0.08$ $\mu$m, $\omega_\mathrm{rad} = 2\pi \times 193$ THz, and $\kappa_\mathrm{rad} = 2\pi \times 0.4$ THz. A plane wave is injected perpendicular to the $z$ axis, and two flux monitors are placed far away from the dipole to determine the transmission and reflection coefficients [Fig. \ref{fig:sim_array}(a)]. For the anisotropic dipole, both the dipole and the source are aligned with the $x$ direction. For the isotropic dipole, the source polarization is rotated by $30^\circ$ in the $xy$ plane to also demonstrate the isotropic polarizability of the dipole.

The simulated transmission and reflection spectra are presented in Fig. \ref{fig:sim_array}(b). For both anisotropic and isotropic dipole implementations, the spectra have a Lorentzian profile and demonstrate the perfect reflection of incident on-resonance waves. The spectral linewidth can be expressed as $3(\lambda/d)^2/(4\pi) \times \kappa_\mathrm{rad}$ with $d$ the dipole spacing \cite{shahmoon2017cooperative}, and is in agreement with the simulated values within $1\%$. The transmission and reflection coefficients sum to $1$ since no energy is lost to the dipole.

To further illustrate the role of intrinsic losses (or lack thereof), we also performed simulations where the intrinsic loss of the dipole is set to be equal to the radiation loss in free space [Fig. \ref{fig:sim_array}(c)]. In this case, the maximum reflection coefficient is reduced and the sum of transmission and reflection coefficients is smaller than $1$. These results further highlight the necessity of self-consistent implementations of dipole radiation losses.

\subsection{Dipoles near a mirror}

\begin{figure}
\centering
\includegraphics[width=132mm]{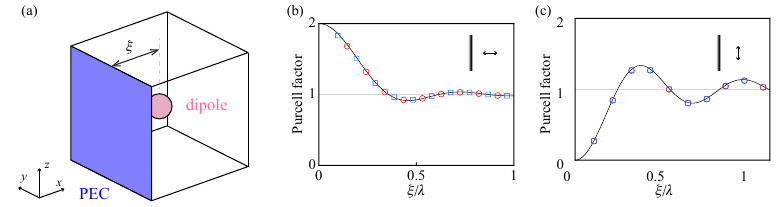}
\caption{Simulation of dipole arrays.
(a) Simulation setup (not to scale). A complete geometric description of the simulation domain can be found in the main text.
(b-c) Theoretical (black curve) and simulated (red circles for anisotropic dipoles, blue squares for isotropic dipoles) Purcell factors when the dipole is polarized perpendicular and parallel to the mirror, respectively.
}
\label{fig:sim_mirror}
\end{figure}

The dipole implementations are also capable of directly simulating the modification of radiation rates in the presence of complex electromagnetic environments. To this end, we simulate the Purcell factors when the dipole is placed near an ideal mirror. Due to interference from the dipole and its mirror image, the radiation rate of a point dipole oscillates with respect to the distance from the mirror \cite{drexhage1970influence}.
In the FDTD setup, we assume a perfect electric conductor (PEC) boundary condition on the $-x$ boundary, and its distance from the dipole $\xi$ is varied between simulations [Fig. \ref{fig:sim_mirror}(a)]. To suppress quadrupole excitations from the isotropic dipole, the grid length has been reduced to $a = 0.05$ $\mu$m which is $\sim \lambda/30$ ($\alpha \approx 0.20$). The radiation rate in free space for both types of dipoles has been set to $0.2$ THz. A TFSF source is used to excite the dipole, where the incident direction is in the $xy$ plane toward the mirror with an incident angle of $45^\circ$ and the light is p-polarized. The radiation rate $\kappa_\mathrm{rad}$ can then be extracted from the exponentially decaying part of the time-dependent field. For the isotropic dipole, the dipole modes parallel and perpendicular to the mirror are excited simultaneously, and their field profiles can be separated by considering their symmetry properties (e.g. mirror symmetry with respect to the $xz$ plane). In contrast, the anisotropic dipole can only be polarized in a single direction, and two different setups are required to characterize the decay rates.
The simulated Purcell factors are presented in Figs. \ref{fig:sim_mirror}(b-c). Both anisotropic and isotropic dipole implementations match theoretical predictions \cite{drexhage1970influence} (without free parameters) with high precision.

\section{Conclusions} \label{sec:conclusion}

Using a Hamiltonian formalism, we explicitly showed that radiative losses do not contribute to the phenomenological damping term in the Lorentz-Drude model if the driving field is interpreted as the total field at the dipole. The resulting dynamics is equivalent to including radiative damping contributions in the damping term while interpreting the driving field as the external field. Apart from analytical examples including the complete reflection effects, the formalism also has direct applications in FDTD simulations, where dipoles in the radiative limit are represented as lossless Lorentz-Drude mediums and the total field at the dipole can be obtained directly. In addition to validating the Hamiltonian formalism, the FDTD simulations are capable of directly reproducing the scattering coefficient and the dipole radiation rate in various electromagnetic environments. This is in contrast to previous methods where the radiation rate is inferred from computations of the Purcell enhancement factor based on the radiated power \cite{novotny2006principles}.

This work is supported by a MURI project from the U.S. Air Force Office of Scientific Research (Grant No. FA9550-21-1-0312).

\bibliography{main}

\begin{thebibliography}{30}%
\makeatletter
\providecommand \@ifxundefined [1]{%
 \@ifx{#1\undefined}
}%
\providecommand \@ifnum [1]{%
 \ifnum #1\expandafter \@firstoftwo
 \else \expandafter \@secondoftwo
 \fi
}%
\providecommand \@ifx [1]{%
 \ifx #1\expandafter \@firstoftwo
 \else \expandafter \@secondoftwo
 \fi
}%
\providecommand \natexlab [1]{#1}%
\providecommand \enquote  [1]{``#1''}%
\providecommand \bibnamefont  [1]{#1}%
\providecommand \bibfnamefont [1]{#1}%
\providecommand \citenamefont [1]{#1}%
\providecommand \href@noop [0]{\@secondoftwo}%
\providecommand \href [0]{\begingroup \@sanitize@url \@href}%
\providecommand \@href[1]{\@@startlink{#1}\@@href}%
\providecommand \@@href[1]{\endgroup#1\@@endlink}%
\providecommand \@sanitize@url [0]{\catcode `\\12\catcode `\$12\catcode `\&12\catcode `\#12\catcode `\^12\catcode `\_12\catcode `\%12\relax}%
\providecommand \@@startlink[1]{}%
\providecommand \@@endlink[0]{}%
\providecommand \url  [0]{\begingroup\@sanitize@url \@url }%
\providecommand \@url [1]{\endgroup\@href {#1}{\urlprefix }}%
\providecommand \urlprefix  [0]{URL }%
\providecommand \Eprint [0]{\href }%
\providecommand \doibase [0]{https://doi.org/}%
\providecommand \selectlanguage [0]{\@gobble}%
\providecommand \bibinfo  [0]{\@secondoftwo}%
\providecommand \bibfield  [0]{\@secondoftwo}%
\providecommand \translation [1]{[#1]}%
\providecommand \BibitemOpen [0]{}%
\providecommand \bibitemStop [0]{}%
\providecommand \bibitemNoStop [0]{.\EOS\space}%
\providecommand \EOS [0]{\spacefactor3000\relax}%
\providecommand \BibitemShut  [1]{\csname bibitem#1\endcsname}%
\let\auto@bib@innerbib\@empty
\bibitem [{\citenamefont {Lorentz}(1916)}]{lorentz1916theory}%
  \BibitemOpen
  \bibfield  {author} {\bibinfo {author} {\bibfnamefont {H.~A.}\ \bibnamefont {Lorentz}},\ }\href@noop {} {\emph {\bibinfo {title} {The {Theory} of {Electrons} and {Its} {Applications} to the {Phenomena} of {Light} and {Radiant} {Heat}: {A} {Course} of {Lectures} {Delivered} in {Columbia} {University}, {New} {York} in {March} and {April}, 1906}}},\ Vol.~\bibinfo {volume} {29}\ (\bibinfo  {publisher} {Teubner},\ \bibinfo {year} {1916})\BibitemShut {NoStop}%
\bibitem [{\citenamefont {Bohren}\ and\ \citenamefont {Huffman}(2004)}]{bohren2004absorption}%
  \BibitemOpen
  \bibfield  {author} {\bibinfo {author} {\bibfnamefont {C.~F.}\ \bibnamefont {Bohren}}\ and\ \bibinfo {author} {\bibfnamefont {D.~R.}\ \bibnamefont {Huffman}},\ }\href {https://doi.org/10.1002/9783527618156} {\emph {\bibinfo {title} {Absorption and scattering of light by small particles}}}\ (\bibinfo  {publisher} {Wiley-CH Verlag GmbH \& Co KGaA},\ \bibinfo {address} {New York},\ \bibinfo {year} {2004})\BibitemShut {NoStop}%
\bibitem [{\citenamefont {Rakić}\ \emph {et~al.}(1998)\citenamefont {Rakić}, \citenamefont {Djurišić}, \citenamefont {Elazar},\ and\ \citenamefont {Majewski}}]{rakic1998optical}%
  \BibitemOpen
  \bibfield  {author} {\bibinfo {author} {\bibfnamefont {A.~D.}\ \bibnamefont {Rakić}}, \bibinfo {author} {\bibfnamefont {A.~B.}\ \bibnamefont {Djurišić}}, \bibinfo {author} {\bibfnamefont {J.~M.}\ \bibnamefont {Elazar}},\ and\ \bibinfo {author} {\bibfnamefont {M.~L.}\ \bibnamefont {Majewski}},\ }\bibfield  {title} {\bibinfo {title} {Optical properties of metallic films for vertical-cavity optoelectronic devices},\ }\href {https://doi.org/10.1364/AO.37.005271} {\bibfield  {journal} {\bibinfo  {journal} {Applied Optics}\ }\textbf {\bibinfo {volume} {37}},\ \bibinfo {pages} {5271} (\bibinfo {year} {1998})}\BibitemShut {NoStop}%
\bibitem [{\citenamefont {Vial}\ \emph {et~al.}(2005)\citenamefont {Vial}, \citenamefont {Grimault}, \citenamefont {Macías}, \citenamefont {Barchiesi},\ and\ \citenamefont {De~La~Chapelle}}]{vial2005improved}%
  \BibitemOpen
  \bibfield  {author} {\bibinfo {author} {\bibfnamefont {A.}~\bibnamefont {Vial}}, \bibinfo {author} {\bibfnamefont {A.-S.}\ \bibnamefont {Grimault}}, \bibinfo {author} {\bibfnamefont {D.}~\bibnamefont {Macías}}, \bibinfo {author} {\bibfnamefont {D.}~\bibnamefont {Barchiesi}},\ and\ \bibinfo {author} {\bibfnamefont {M.~L.}\ \bibnamefont {De~La~Chapelle}},\ }\bibfield  {title} {\bibinfo {title} {Improved analytical fit of gold dispersion: Application to the modeling of extinction spectra with a finite-difference time-domain method},\ }\href {https://doi.org/10.1103/PhysRevB.71.085416} {\bibfield  {journal} {\bibinfo  {journal} {Physical Review B}\ }\textbf {\bibinfo {volume} {71}},\ \bibinfo {pages} {085416} (\bibinfo {year} {2005})}\BibitemShut {NoStop}%
\bibitem [{\citenamefont {Siegman}(1986)}]{siegman1986lasers}%
  \BibitemOpen
  \bibfield  {author} {\bibinfo {author} {\bibfnamefont {A.~E.}\ \bibnamefont {Siegman}},\ }\href@noop {} {\emph {\bibinfo {title} {Lasers}}}\ (\bibinfo  {publisher} {University Science Books},\ \bibinfo {address} {Mill Valley, California},\ \bibinfo {year} {1986})\BibitemShut {NoStop}%
\bibitem [{\citenamefont {Shen}\ and\ \citenamefont {Fan}(2005)}]{shen2005coherent}%
  \BibitemOpen
  \bibfield  {author} {\bibinfo {author} {\bibfnamefont {J.~T.}\ \bibnamefont {Shen}}\ and\ \bibinfo {author} {\bibfnamefont {S.}~\bibnamefont {Fan}},\ }\bibfield  {title} {\bibinfo {title} {Coherent photon transport from spontaneous emission in one-dimensional waveguides},\ }\href {https://doi.org/10.1364/OL.30.002001} {\bibfield  {journal} {\bibinfo  {journal} {Optics Letters}\ }\textbf {\bibinfo {volume} {30}},\ \bibinfo {pages} {2001} (\bibinfo {year} {2005})}\BibitemShut {NoStop}%
\bibitem [{\citenamefont {Chang}\ \emph {et~al.}(2007)\citenamefont {Chang}, \citenamefont {Sørensen}, \citenamefont {Demler},\ and\ \citenamefont {Lukin}}]{chang2007single-photon}%
  \BibitemOpen
  \bibfield  {author} {\bibinfo {author} {\bibfnamefont {D.~E.}\ \bibnamefont {Chang}}, \bibinfo {author} {\bibfnamefont {A.~S.}\ \bibnamefont {Sørensen}}, \bibinfo {author} {\bibfnamefont {E.~A.}\ \bibnamefont {Demler}},\ and\ \bibinfo {author} {\bibfnamefont {M.~D.}\ \bibnamefont {Lukin}},\ }\bibfield  {title} {\bibinfo {title} {A single-photon transistor using nanoscale surface plasmons},\ }\href {https://doi.org/10.1038/nphys708} {\bibfield  {journal} {\bibinfo  {journal} {Nature Physics}\ }\textbf {\bibinfo {volume} {3}},\ \bibinfo {pages} {807} (\bibinfo {year} {2007})}\BibitemShut {NoStop}%
\bibitem [{\citenamefont {Lodahl}\ \emph {et~al.}(2015)\citenamefont {Lodahl}, \citenamefont {Mahmoodian},\ and\ \citenamefont {Stobbe}}]{lodahl2015interfacing}%
  \BibitemOpen
  \bibfield  {author} {\bibinfo {author} {\bibfnamefont {P.}~\bibnamefont {Lodahl}}, \bibinfo {author} {\bibfnamefont {S.}~\bibnamefont {Mahmoodian}},\ and\ \bibinfo {author} {\bibfnamefont {S.}~\bibnamefont {Stobbe}},\ }\bibfield  {title} {\bibinfo {title} {Interfacing single photons and single quantum dots with photonic nanostructures},\ }\href {https://doi.org/10.1103/RevModPhys.87.347} {\bibfield  {journal} {\bibinfo  {journal} {Reviews of Modern Physics}\ }\textbf {\bibinfo {volume} {87}},\ \bibinfo {pages} {347} (\bibinfo {year} {2015})}\BibitemShut {NoStop}%
\bibitem [{\citenamefont {Bettles}\ \emph {et~al.}(2016)\citenamefont {Bettles}, \citenamefont {Gardiner},\ and\ \citenamefont {Adams}}]{bettles2016enhanced}%
  \BibitemOpen
  \bibfield  {author} {\bibinfo {author} {\bibfnamefont {R.~J.}\ \bibnamefont {Bettles}}, \bibinfo {author} {\bibfnamefont {S.~A.}\ \bibnamefont {Gardiner}},\ and\ \bibinfo {author} {\bibfnamefont {C.~S.}\ \bibnamefont {Adams}},\ }\bibfield  {title} {\bibinfo {title} {Enhanced optical cross section via collective coupling of atomic dipoles in a 2d array},\ }\href {https://doi.org/10.1103/PhysRevLett.116.103602} {\bibfield  {journal} {\bibinfo  {journal} {Physical Review Letters}\ }\textbf {\bibinfo {volume} {116}},\ \bibinfo {pages} {103602} (\bibinfo {year} {2016})}\BibitemShut {NoStop}%
\bibitem [{\citenamefont {Shahmoon}\ \emph {et~al.}(2017)\citenamefont {Shahmoon}, \citenamefont {Wild}, \citenamefont {Lukin},\ and\ \citenamefont {Yelin}}]{shahmoon2017cooperative}%
  \BibitemOpen
  \bibfield  {author} {\bibinfo {author} {\bibfnamefont {E.}~\bibnamefont {Shahmoon}}, \bibinfo {author} {\bibfnamefont {D.~S.}\ \bibnamefont {Wild}}, \bibinfo {author} {\bibfnamefont {M.~D.}\ \bibnamefont {Lukin}},\ and\ \bibinfo {author} {\bibfnamefont {S.~F.}\ \bibnamefont {Yelin}},\ }\bibfield  {title} {\bibinfo {title} {Cooperative resonances in light scattering from two-dimensional atomic arrays},\ }\href {https://doi.org/10.1103/PhysRevLett.118.113601} {\bibfield  {journal} {\bibinfo  {journal} {Physical Review Letters}\ }\textbf {\bibinfo {volume} {118}},\ \bibinfo {pages} {113601} (\bibinfo {year} {2017})}\BibitemShut {NoStop}%
\bibitem [{\citenamefont {Rui}\ \emph {et~al.}(2020)\citenamefont {Rui}, \citenamefont {Wei}, \citenamefont {Rubio-Abadal}, \citenamefont {Hollerith}, \citenamefont {Zeiher}, \citenamefont {Stamper-Kurn}, \citenamefont {Gross},\ and\ \citenamefont {Bloch}}]{rui2020subradiant}%
  \BibitemOpen
  \bibfield  {author} {\bibinfo {author} {\bibfnamefont {J.}~\bibnamefont {Rui}}, \bibinfo {author} {\bibfnamefont {D.}~\bibnamefont {Wei}}, \bibinfo {author} {\bibfnamefont {A.}~\bibnamefont {Rubio-Abadal}}, \bibinfo {author} {\bibfnamefont {S.}~\bibnamefont {Hollerith}}, \bibinfo {author} {\bibfnamefont {J.}~\bibnamefont {Zeiher}}, \bibinfo {author} {\bibfnamefont {D.~M.}\ \bibnamefont {Stamper-Kurn}}, \bibinfo {author} {\bibfnamefont {C.}~\bibnamefont {Gross}},\ and\ \bibinfo {author} {\bibfnamefont {I.}~\bibnamefont {Bloch}},\ }\bibfield  {title} {\bibinfo {title} {A subradiant optical mirror formed by a single structured atomic layer},\ }\href {https://doi.org/10.1038/s41586-020-2463-x} {\bibfield  {journal} {\bibinfo  {journal} {Nature}\ }\textbf {\bibinfo {volume} {583}},\ \bibinfo {pages} {369} (\bibinfo {year} {2020})}\BibitemShut {NoStop}%
\bibitem [{\citenamefont {Sheremet}\ \emph {et~al.}(2023)\citenamefont {Sheremet}, \citenamefont {Petrov}, \citenamefont {Iorsh}, \citenamefont {Poshakinskiy},\ and\ \citenamefont {Poddubny}}]{sheremet2023waveguide}%
  \BibitemOpen
  \bibfield  {author} {\bibinfo {author} {\bibfnamefont {A.~S.}\ \bibnamefont {Sheremet}}, \bibinfo {author} {\bibfnamefont {M.~I.}\ \bibnamefont {Petrov}}, \bibinfo {author} {\bibfnamefont {I.~V.}\ \bibnamefont {Iorsh}}, \bibinfo {author} {\bibfnamefont {A.~V.}\ \bibnamefont {Poshakinskiy}},\ and\ \bibinfo {author} {\bibfnamefont {A.~N.}\ \bibnamefont {Poddubny}},\ }\bibfield  {title} {\bibinfo {title} {Waveguide quantum electrodynamics: Collective radiance and photon-photon correlations},\ }\href {https://doi.org/10.1103/RevModPhys.95.015002} {\bibfield  {journal} {\bibinfo  {journal} {Reviews of Modern Physics}\ }\textbf {\bibinfo {volume} {95}},\ \bibinfo {pages} {015002} (\bibinfo {year} {2023})}\BibitemShut {NoStop}%
\bibitem [{\citenamefont {Stobińska}\ \emph {et~al.}(2009)\citenamefont {Stobińska}, \citenamefont {Alber},\ and\ \citenamefont {Leuchs}}]{stobinska2009perfect}%
  \BibitemOpen
  \bibfield  {author} {\bibinfo {author} {\bibfnamefont {M.}~\bibnamefont {Stobińska}}, \bibinfo {author} {\bibfnamefont {G.}~\bibnamefont {Alber}},\ and\ \bibinfo {author} {\bibfnamefont {G.}~\bibnamefont {Leuchs}},\ }\bibfield  {title} {\bibinfo {title} {Perfect excitation of a matter qubit by a single photon in free space},\ }\href {https://doi.org/10.1209/0295-5075/86/14007} {\bibfield  {journal} {\bibinfo  {journal} {EPL (Europhysics Letters)}\ }\textbf {\bibinfo {volume} {86}},\ \bibinfo {pages} {14007} (\bibinfo {year} {2009})}\BibitemShut {NoStop}%
\bibitem [{\citenamefont {Schelew}\ \emph {et~al.}(2017)\citenamefont {Schelew}, \citenamefont {Ge}, \citenamefont {Hughes}, \citenamefont {Pond},\ and\ \citenamefont {Young}}]{schelew2017self-consistent}%
  \BibitemOpen
  \bibfield  {author} {\bibinfo {author} {\bibfnamefont {E.}~\bibnamefont {Schelew}}, \bibinfo {author} {\bibfnamefont {R.-C.}\ \bibnamefont {Ge}}, \bibinfo {author} {\bibfnamefont {S.}~\bibnamefont {Hughes}}, \bibinfo {author} {\bibfnamefont {J.}~\bibnamefont {Pond}},\ and\ \bibinfo {author} {\bibfnamefont {J.~F.}\ \bibnamefont {Young}},\ }\bibfield  {title} {\bibinfo {title} {Self-consistent numerical modeling of radiatively damped lorentz oscillators},\ }\href {https://doi.org/10.1103/PhysRevA.95.063853} {\bibfield  {journal} {\bibinfo  {journal} {Physical Review A}\ }\textbf {\bibinfo {volume} {95}},\ \bibinfo {pages} {063853} (\bibinfo {year} {2017})}\BibitemShut {NoStop}%
\bibitem [{\citenamefont {Zhou}\ \emph {et~al.}(2024)\citenamefont {Zhou}, \citenamefont {Gangaraj}, \citenamefont {Zhou},\ and\ \citenamefont {Yu}}]{zhou2024simulating}%
  \BibitemOpen
  \bibfield  {author} {\bibinfo {author} {\bibfnamefont {Q.}~\bibnamefont {Zhou}}, \bibinfo {author} {\bibfnamefont {S.}~\bibnamefont {Gangaraj}}, \bibinfo {author} {\bibfnamefont {M.}~\bibnamefont {Zhou}},\ and\ \bibinfo {author} {\bibfnamefont {Z.}~\bibnamefont {Yu}},\ }\bibfield  {title} {\bibinfo {title} {Simulating quantum emitters in arbitrary photonic environments using fdtd: beyond the semi-classical regime},\ }\href@noop {} {\bibfield  {journal} {\bibinfo  {journal} {arXiv:2410.16118}\ } (\bibinfo {year} {2024})}\BibitemShut {NoStop}%
\bibitem [{\citenamefont {Landau}\ and\ \citenamefont {Lifshitz}(2010)}]{landau2010classical}%
  \BibitemOpen
  \bibfield  {author} {\bibinfo {author} {\bibfnamefont {L.~D.}\ \bibnamefont {Landau}}\ and\ \bibinfo {author} {\bibfnamefont {E.~M.}\ \bibnamefont {Lifshitz}},\ }\href@noop {} {\emph {\bibinfo {title} {The classical theory of fields}}}\ (\bibinfo  {publisher} {Elsevier},\ \bibinfo {year} {2010})\BibitemShut {NoStop}%
\bibitem [{\citenamefont {Jackson}(2002)}]{jackson2002lorenz}%
  \BibitemOpen
  \bibfield  {author} {\bibinfo {author} {\bibfnamefont {J.~D.}\ \bibnamefont {Jackson}},\ }\bibfield  {title} {\bibinfo {title} {From lorenz to coulomb and other explicit gauge transformations},\ }\href {https://doi.org/10.1119/1.1491265} {\bibfield  {journal} {\bibinfo  {journal} {American Journal of Physics}\ }\textbf {\bibinfo {volume} {70}},\ \bibinfo {pages} {917} (\bibinfo {year} {2002})}\BibitemShut {NoStop}%
\bibitem [{\citenamefont {Raman}\ and\ \citenamefont {Fan}(2010)}]{raman2010photonic}%
  \BibitemOpen
  \bibfield  {author} {\bibinfo {author} {\bibfnamefont {A.}~\bibnamefont {Raman}}\ and\ \bibinfo {author} {\bibfnamefont {S.}~\bibnamefont {Fan}},\ }\bibfield  {title} {\bibinfo {title} {Photonic band structure of dispersive metamaterials formulated as a {H}ermitian eigenvalue problem},\ }\href {https://doi.org/10.1103/PhysRevLett.104.087401} {\bibfield  {journal} {\bibinfo  {journal} {Physical Review Letters}\ }\textbf {\bibinfo {volume} {104}},\ \bibinfo {pages} {087401} (\bibinfo {year} {2010})}\BibitemShut {NoStop}%
\bibitem [{\citenamefont {Jackson}(2009)}]{jackson2009classical}%
  \BibitemOpen
  \bibfield  {author} {\bibinfo {author} {\bibfnamefont {J.~D.}\ \bibnamefont {Jackson}},\ }\href@noop {} {\emph {\bibinfo {title} {Classical electrodynamics}}}\ (\bibinfo  {publisher} {Wiley},\ \bibinfo {address} {NY},\ \bibinfo {year} {2009})\BibitemShut {NoStop}%
\bibitem [{\citenamefont {De~Vries}\ \emph {et~al.}(1998)\citenamefont {De~Vries}, \citenamefont {Van~Coevorden},\ and\ \citenamefont {Lagendijk}}]{vries1998point}%
  \BibitemOpen
  \bibfield  {author} {\bibinfo {author} {\bibfnamefont {P.}~\bibnamefont {De~Vries}}, \bibinfo {author} {\bibfnamefont {D.~V.}\ \bibnamefont {Van~Coevorden}},\ and\ \bibinfo {author} {\bibfnamefont {A.}~\bibnamefont {Lagendijk}},\ }\bibfield  {title} {\bibinfo {title} {Point scatterers for classical waves},\ }\href {https://doi.org/10.1103/RevModPhys.70.447} {\bibfield  {journal} {\bibinfo  {journal} {Reviews of Modern Physics}\ }\textbf {\bibinfo {volume} {70}},\ \bibinfo {pages} {447} (\bibinfo {year} {1998})}\BibitemShut {NoStop}%
\bibitem [{\citenamefont {Moskalensky}\ and\ \citenamefont {Yurkin}(2021)}]{moskalensky2021point}%
  \BibitemOpen
  \bibfield  {author} {\bibinfo {author} {\bibfnamefont {A.~E.}\ \bibnamefont {Moskalensky}}\ and\ \bibinfo {author} {\bibfnamefont {M.~A.}\ \bibnamefont {Yurkin}},\ }\bibfield  {title} {\bibinfo {title} {A point electric dipole: From basic optical properties to the fluctuation–dissipation theorem},\ }\href {https://doi.org/10.1016/j.revip.2020.100047} {\bibfield  {journal} {\bibinfo  {journal} {Reviews in Physics}\ }\textbf {\bibinfo {volume} {6}},\ \bibinfo {pages} {100047} (\bibinfo {year} {2021})}\BibitemShut {NoStop}%
\bibitem [{\citenamefont {Gardiner}\ and\ \citenamefont {Collett}(1985)}]{gardiner1985input}%
  \BibitemOpen
  \bibfield  {author} {\bibinfo {author} {\bibfnamefont {C.~W.}\ \bibnamefont {Gardiner}}\ and\ \bibinfo {author} {\bibfnamefont {M.~J.}\ \bibnamefont {Collett}},\ }\bibfield  {title} {\bibinfo {title} {Input and output in damped quantum systems: Quantum stochastic differential equations and the master equation},\ }\href {https://doi.org/10.1103/PhysRevA.31.3761} {\bibfield  {journal} {\bibinfo  {journal} {Physical Review A}\ }\textbf {\bibinfo {volume} {31}},\ \bibinfo {pages} {3761} (\bibinfo {year} {1985})}\BibitemShut {NoStop}%
\bibitem [{\citenamefont {Gardiner}\ and\ \citenamefont {Zoller}(2000)}]{gardiner2000quantum}%
  \BibitemOpen
  \bibfield  {author} {\bibinfo {author} {\bibfnamefont {C.~W.}\ \bibnamefont {Gardiner}}\ and\ \bibinfo {author} {\bibfnamefont {P.}~\bibnamefont {Zoller}},\ }\href@noop {} {\emph {\bibinfo {title} {Quantum noise: a handbook of Markovian and non-Markovian quantum stochastic methods with applications to quantum optics}}}\ (\bibinfo  {publisher} {Springer},\ \bibinfo {address} {Berlin Heidelberg},\ \bibinfo {year} {2000})\BibitemShut {NoStop}%
\bibitem [{\citenamefont {Fan}\ \emph {et~al.}(2010)\citenamefont {Fan}, \citenamefont {Kocaba{\c{s}}},\ and\ \citenamefont {Shen}}]{fan2010input-output}%
  \BibitemOpen
  \bibfield  {author} {\bibinfo {author} {\bibfnamefont {S.}~\bibnamefont {Fan}}, \bibinfo {author} {\bibfnamefont {{\c{S}}.~E.}\ \bibnamefont {Kocaba{\c{s}}}},\ and\ \bibinfo {author} {\bibfnamefont {J.-T.}\ \bibnamefont {Shen}},\ }\bibfield  {title} {\bibinfo {title} {Input-output formalism for few-photon transport in one-dimensional nanophotonic waveguides coupled to a qubit},\ }\href {https://doi.org/10.1103/PhysRevA.82.063821} {\bibfield  {journal} {\bibinfo  {journal} {Physical Review A}\ }\textbf {\bibinfo {volume} {82}},\ \bibinfo {pages} {063821} (\bibinfo {year} {2010})}\BibitemShut {NoStop}%
\bibitem [{\citenamefont {Taflove}\ and\ \citenamefont {Hagness}(2010)}]{taflove2010computational}%
  \BibitemOpen
  \bibfield  {author} {\bibinfo {author} {\bibfnamefont {A.}~\bibnamefont {Taflove}}\ and\ \bibinfo {author} {\bibfnamefont {S.~C.}\ \bibnamefont {Hagness}},\ }\href@noop {} {\emph {\bibinfo {title} {Computational electrodynamics: the finite-difference time-domain method}}}\ (\bibinfo  {publisher} {Artech House},\ \bibinfo {address} {Boston, Mass.},\ \bibinfo {year} {2010})\BibitemShut {NoStop}%
\bibitem [{\citenamefont {Ma}\ \emph {et~al.}(2005)\citenamefont {Ma}, \citenamefont {Rayner},\ and\ \citenamefont {Parini}}]{ma2005discrete}%
  \BibitemOpen
  \bibfield  {author} {\bibinfo {author} {\bibfnamefont {W.}~\bibnamefont {Ma}}, \bibinfo {author} {\bibfnamefont {M.}~\bibnamefont {Rayner}},\ and\ \bibinfo {author} {\bibfnamefont {C.}~\bibnamefont {Parini}},\ }\bibfield  {title} {\bibinfo {title} {Discrete green's function formulation of the fdtd method and its application in antenna modeling},\ }\href {https://doi.org/10.1109/TAP.2004.838797} {\bibfield  {journal} {\bibinfo  {journal} {IEEE Transactions on Antennas and Propagation}\ }\textbf {\bibinfo {volume} {53}},\ \bibinfo {pages} {339} (\bibinfo {year} {2005})}\BibitemShut {NoStop}%
\bibitem [{\citenamefont {Kastner}(2006)}]{kastner2006multidimensional}%
  \BibitemOpen
  \bibfield  {author} {\bibinfo {author} {\bibfnamefont {R.}~\bibnamefont {Kastner}},\ }\bibfield  {title} {\bibinfo {title} {A multidimensional $z$-transform evaluation of the discrete finite difference time domain green's function},\ }\href {https://doi.org/10.1109/TAP.2006.872674} {\bibfield  {journal} {\bibinfo  {journal} {IEEE Transactions on Antennas and Propagation}\ }\textbf {\bibinfo {volume} {54}},\ \bibinfo {pages} {1215} (\bibinfo {year} {2006})}\BibitemShut {NoStop}%
\bibitem [{tid(2025)}]{tidy3d}%
  \BibitemOpen
  \href {https://www.flexcompute.com/tidy3d/solver} {\bibinfo {title} {Python-driven fdtd software: Tidy3d $\vert$ flexcompute}} (\bibinfo {year} {2025})\BibitemShut {NoStop}%
\bibitem [{\citenamefont {Drexhage}(1970)}]{drexhage1970influence}%
  \BibitemOpen
  \bibfield  {author} {\bibinfo {author} {\bibfnamefont {K.}~\bibnamefont {Drexhage}},\ }\bibfield  {title} {\bibinfo {title} {Influence of a dielectric interface on fluorescence decay time},\ }\href {https://doi.org/10.1016/0022-2313(70)90082-7} {\bibfield  {journal} {\bibinfo  {journal} {Journal of Luminescence}\ }\textbf {\bibinfo {volume} {1}},\ \bibinfo {pages} {693} (\bibinfo {year} {1970})}\BibitemShut {NoStop}%
\bibitem [{\citenamefont {Novotny}\ and\ \citenamefont {Hecht}(2006)}]{novotny2006principles}%
  \BibitemOpen
  \bibfield  {author} {\bibinfo {author} {\bibfnamefont {L.}~\bibnamefont {Novotny}}\ and\ \bibinfo {author} {\bibfnamefont {B.}~\bibnamefont {Hecht}},\ }\href {https://doi.org/10.1017/CBO9780511813535} {\emph {\bibinfo {title} {Principles of Nano-Optics}}}\ (\bibinfo  {publisher} {Cambridge University Press},\ \bibinfo {year} {2006})\BibitemShut {NoStop}%
\end{thebibliography}%

\end{document}